\documentstyle[12pt]{article}
\newcommand\beq{\begin{equation}}
\newcommand\eeq{\end{equation}}
\newcommand\bea{\begin{eqnarray}}
\newcommand\eea{\end{eqnarray}}

\begin{document}
\vspace{-2.0cm}
\bigskip

\centerline{\Large \bf Coherent States For $SU(3)$} 
\vskip .8 true cm

\begin{center} 
{\bf Manu Mathur} \footnote{manu@boson.bose.res.in} and 
{\bf Diptiman Sen} \footnote{diptiman@cts.iisc.ernet.in}  
\vskip 1.0 true cm

{$^1$}S. N. Bose National Centre for Basic Sciences \\ 
JD Block, Sector III, Salt Lake City, Calcutta 700091, India

\vskip .5 true cm

$^2$ Centre for Theoretical Studies  \\  
Indian Institute of Science, Bangalore 560012, India 

\end{center} 
\bigskip

\centerline{\bf Abstract}

We define coherent states for $SU(3)$ using six bosonic creation and 
annihilation operators. These coherent states are explicitly characterized by 
six complex numbers with constraints. For the completely symmetric 
representations $(n,0)$ and $(0,m)$, only three of the bosonic operators are 
required. For mixed representations $(n,m)$, all six operators are required. 
The coherent states provide a resolution of identity, satisfy the continuity 
property, and possess a variety of group theoretic properties. We introduce an 
explicit parameterization of the group $SU(3)$ and the corresponding 
integration measure. Finally, we discuss the path integral formalism for a 
problem in which the Hamiltonian is a function of $SU(3)$ operators at each 
site.

\vskip .4 true cm
\noindent PACS: ~02.20.-a 
\vskip .4 true cm

\section{\bf Introduction}

\indent Coherent states 
have been used for a long time in different areas of physics 
\cite{klauder,perelomov}. In condensed matter physics, coherent states for the 
Lie group $SU(2)$ have been extensively used to study Heisenberg spin systems 
using the path integral formalism \cite{arovas,manous,fradkin,sachdev}. 
These studies have been generalized to systems with $SU(N)$ symmetry; these 
studies have usually been restricted to the completely symmetric 
representations \cite{manous,sun}. However, there is a recent discussion of 
coherent states for arbitrary irreducible representations of $SU(3)$ in Ref. 
\cite{gnutzmann}. The purpose of our work is to 
discuss a coherent state formalism which is valid for all representations of 
$SU(3)$, and to give an explicit characterization of them in terms of complex 
numbers and the states of some harmonic oscillators. (Our work differs in this 
respect from Ref. \cite{gnutzmann} which does not use harmonic oscillator 
operators to define the basis states). As we will see, this way of 
characterization is very similar to those used for the Heisenberg-Weyl 
and $SU(2)$ coherent states. But, there are also certain features (such as 
tracelessness) which are redundant in the simpler case of $SU(2)$. 

As additional motivation for our work, we should mention that there 
have been many other studies of $SU(3)$ in 
the recent mathematical physics literature, including the geometric
phase for three-level systems \cite{khanna} and the study of Clebsch-Gordon 
coefficients and the outer multiplicity problem \cite{prakash}. These studies 
do not use coherent states; however our work is likely to shed new light on 
some of these studies. For instance, we will use two triplets of complex
numbers $z$ and $w$ which are similar to the ones used in \cite{prakash}, 
except that we will normalize the triplets to unity. Similarly, it is 
well-known that the geometric phases in the different representations of 
$SU(2)$ may be obtained by integrating around a closed loop the overlap of two 
coherent states which differ infinitesimally from each other 
\cite{fradkin,sachdev}. In the same way, it should be possible to derive the 
geometric phases for $SU(3)$ representations from the coherent states 
discussed below.

The organization of the paper is as follows. Section 2 will motivate our ideas 
and techniques using two examples which are simpler than the $SU(3)$ group. We 
start with the standard group theoretical definitions of the coherent states 
of the Heisenberg-Weyl and $SU(2)$ groups. We then discuss another way 
of defining $SU(2)$ coherent states using the Schwinger or Holstein-Primakoff
representation of the Lie algebra of $SU(2)$ \cite{schwinger} in terms 
of harmonic oscillator creation and annihilation operators. This definition
is discussed in some detail as it can be extended to the $SU(3)$ group. 
We then establish its equivalence with the standard group theoretical coherent 
state definition \cite{perelomov}. In section 3, we generalize the $SU(2)$ Lie 
algebra in terms of harmonic oscillators to the $SU(3)$ group, and construct 
the irreducible representations of $SU(3)$. We describe the structure of 
$SU(3)$ matrices in an explicit way, and provide an integration measure for 
this $8$-dimensional manifold. In section 4, we use this group structure to 
construct a set of $SU(3)$ coherent states which are explicitly characterized 
by a set of complex numbers which are equivalent to $8$ real variables. We 
prove various identities expected for 
coherent states such as the the resolution of identity and a transformation
from a particular coherent state to the general coherent state. In section 5,
we provide an alternative set of coherent states for $SU(3)$ which require
only $5$ real variables; although these share some of the features
of the coherent states defined in section 4, they have a few limitations
arising from the smaller number of variables used. In section 6, we 
discuss how coherent states can be used to develop a path integral formalism
for problems involving $SU(3)$ variables.

\section{\bf Heisenberg-Weyl and $SU(2)$ Coherent States}

\indent
There are many definitions of coherent states used in the literature. However, 
the most essential ingredients common in all these definitions are the 
continuity and completeness properties \cite{klauder}.

\begin{enumerate} 

\item These are states in a Hilbert space $\cal H$ associated which are 
characterized by a set of continuous variables $\lbrace \vec{z} \rbrace$,
and the coherent states $\vert \vec{z} >$ are strongly continuous functions 
of the labels $\lbrace \vec{z} \rbrace$. 

\item There exists a positive measure $d\mu(\vec{z})$ such that the unit 
operator ${\cal I}$ admits the resolution of identity 
\beq
{\cal I} ~=~ \int ~d\mu(\vec{z}) ~\vert \vec{z} > < \vec{z}\vert ~.
\label{roi1} 
\eeq

\end{enumerate} 

\noindent Given a group $G$, the coherent states in a 
given representation $R$ are functions of $q$ parameters  denoted by 
$\lbrace z_1,z_2,...z_q \rbrace$, and are defined as
\beq
\vert \vec{z} > ~\equiv ~T_R \big(g(\vec{z})\big) ~\vert 0 >_R ~.
\label{gp} 
\eeq
Here $T_{R}\big(g(\vec{z})\big)$ is a group element in 
the representation $R$, and $\vert 0>_{R}$ is a fixed vector belonging to 
$R$. In the simplest example of the Heisenberg-Weyl group, 
the Lie algebra contains three generators. It is defined in terms 
of creation annihilation operators $(a, a^{\dagger})$ satisfying
\beq
[a,a^{\dagger}]= {\cal I}, ~~~[a,{\cal I}] =0, ~~~[a^{\dagger}, {\cal I}]=0 ~.
\eeq
This algebra has only one infinite dimensional irreducible 
representation which 
can be characterized by occupation number states $\vert n> \equiv 
{(a^{\dagger})^n \over \sqrt {n!}} \vert 0>$ with $n=0, 1, 2 ...$.   
A generic group element in (\ref{gp}) can be characterized by 
$T\big(g\big) = \exp ~(i\alpha {\cal I} + z a^{\dagger} - \bar{z}a)$ 
with an angle $\alpha$ and a complex parameter $z$. Therefore, 
\bea
\vert \alpha,z>_\infty ~&=&~ \exp (i\alpha) ~\vert z> , \nonumber \\
\vert z> ~&=&~ \exp (z a^{\dagger} - \bar{z} a) ~\vert 0>  
= \sum_{n=0}^{\infty} F_n (z) ~\vert n> ~,
\label{wcs} 
\eea
where the sum runs over all the basis vectors of the infinite dimensional 
representation, and 
\beq
F_n (z) ~=~ {z^n \over \sqrt{n!}} ~\exp (-|z|^2 /2)
\label{co1} 
\eeq
are the coherent state expansion coefficients. This feature, i.e., an 
expansion of the coherent states in terms of basis vectors of a given 
representation with analytic functions of complex variables ($F_n (z)$) as 
coefficients, will also be present in the case of $SU(2)$ and $SU(3)$ groups. 
It is easy to see that  Eq. (\ref{wcs}) provides a resolution of identity as 
in (\ref{roi1}) with the measure $d\mu(z) = dzd{\bar{z}}$. 

We now briefly review the next simplest example, i.e., the coherent states 
associated with the $SU(2)$ group. The $SU(2)$ Lie 
algebra is given by a set of three angular momentum operators $\lbrace \vec{J} 
\rbrace \equiv \lbrace J_1,J_2,J_3 \rbrace$ or equivalently by $\lbrace J_{+},
J_{-},J_{3} \rbrace$, $(J_{\pm} \equiv J_{1} \pm i~ J_{2})$ satisfying
\beq
[J_{3},J_{\pm}] = \pm J_{\pm}, ~~~~~~ [J_{+},J_{-}] = 2 J_{3} ~.
\label{su2a} 
\eeq

\noindent The $SU(2)$ group has a Casimir operator given by $\vec{J} \cdot 
\vec{J}$, and the different irreducible representations are characterized 
by its eigenvalues $j(j+1)$, where $j$ is an integer or half-odd-integer. 
A given basis vector in representation $j$ is labeled by the eigenvalue $m$ 
of $J_3$ as $\vert j,m>$. We characterize the SU(2) group elements U by the 
Euler angles, i.e, $U(\theta,\phi,\psi) \equiv  exp -i \phi J_3  exp -i 
\theta J_2 exp -i \psi J_3$.   The standard group theoretical definition 
(\ref{gp}) takes $\vert 0>_{j}$ in (\ref{gp}) to be the highest 
weight state $\vert j, j>$ and is of the form:
\bea
\vert \hat{n}(\theta,\phi) >_{j} ~&=&~ U(\theta,\phi,\psi) ~\vert j, j> ~, 
\nonumber \\
&=&~ \sum_{m=-j}^{+j} C_{m}(\theta,\phi)  ~\vert j, m> ~, 
\label{st}    
\eea

\noindent In (\ref{st}), the coefficients $C_{m}(\theta,\phi)$ are given by, 
\bea
C_{m}(\theta,\phi) = e^{-i m \phi} \big[{ 2j! \over (j+m)!(j-m)!}\big]^{1 
\over 2} \big[sin {\theta \over 2}\big]^{j-m} \big[cos {\theta \over 2}
\big]^{j+m}   
\label{coeff} 
\eea

\noindent where we have ignored possible phase factors. 

\noindent The algebra in Eq. (\ref{su2a}) can also be realized in terms of 
a doublet of harmonic oscillator creation and annihilation operators 
$\vec{a} \equiv (a_1, a_2)$ and $ \vec{a}^{\dagger} \equiv (a^{\dagger}_1, 
a^{\dagger}_2)$ respectively \cite{schwinger}. 
They satisfy the simpler bosonic commutation relation
$[a_i,a^{\dagger}_j] = \delta_{ij}$ with $i,j =1,2$. The vacuum state 
is $\vert 0,0 >$. In terms of these operators, 
\beq
J^a ~\equiv ~\frac{1}{2} ~a^{\dagger}_i ~(\sigma^a )_{ij} ~a_j ~,
\label{sch} 
\eeq
where $\sigma^a$ denote the Pauli matrices. (We will generally use the 
convention that repeated indices are summed over). It is easy to check that 
the operators in (\ref{sch}) satisfy the $SU(2)$ Lie algebra with the Casimir 
$\vec{J} \cdot \vec{J} \equiv {1 \over 4} \vec{a}^{\dagger} \cdot \vec{a} 
(\vec{a}^{\dagger} \cdot \vec{a} + 2)$. Thus the representations of $SU(2)$ 
can be characterized by the eigenvalues of the occupation number operator; the 
spin value $j$ is equal to $(N_1+ N_2)/2$ where $N_1$ and $N_2$ are the 
eigenvalues of $a^{\dagger}_1 a_1$ and $a^{\dagger}_2 a_2$ respectively. 

With these harmonic oscillator
creation and annihilation operators, another definition of $SU(2)$ coherent 
states is obtained by directly generalizing (\ref{wcs}). We define a doublet 
of complex numbers $(z_1, z_2)$ with the constraint $|z_1|^2 + |z_2|^2 
=1$; this gives $3$ independent real parameters which define the sphere
$S^3$. Let us parameterize 
\beq 
z_1 ~=~ \cos \chi ~e^{i\beta_1} ~, \quad {\rm and} \quad z_2 ~=~ \sin \chi ~
e^{i\beta_2} ~,
\label{z1z2}
\eeq
where $0 \le \chi \le \pi /2$ and $0 \le \beta_1 , \beta_2 < 2\pi$. 
The the integration measure on this space takes the form
\beq
d\Omega_{S^3} ~=~ \frac{1}{\pi^2} ~dz_1 ~d{\bar z}_1 ~dz_2 ~d{\bar z}_2 ~
\delta (|z_1|^2+|z_2|^2-1) ~=~ \frac{1}{2\pi^2} ~\cos \chi ~\sin \chi ~
d\chi ~d\beta_1 ~d\beta_2 ~,
\label{ints3}
\eeq
where we have introduced a normalization factor so that $\int d\Omega_{S^3} 
=1$. The $SU(2)$ coherent state in the representation $N$ is now defined as
\bea
\vert z_1, z_2>_{N=2j} & ~=~ &  \delta_{\vec{a}^{\dagger} \cdot \vec{a},N} 
~\sqrt{N!} ~
\exp \Big( \vec{z} \cdot \vec{a}^\dagger \Big) ~\vert 0,0> \nonumber \\
& =~ & \sum_{N_1,N_2} \hspace{-0.2cm} {}^{\prime} ~F_{N_1,N_2} ~\vert N_1, 
N_2>_j ~.
\label{cs2} 
\eea
In the second equation above, 
the $\sum {}^\prime$ implies that only the terms satisfying the constraint 
$a^\dagger \cdot a = N \equiv 2j$ are included or equivalently that
\beq
N_1 + N_2 = N ~.
\label{con2} 
\eeq
With (\ref{con2}), the states $\vert N_1, N_2>_j$ form a $(2j+1)$-dimensional 
representation of $SU(2)$. The expansion coefficients $F_{N_1,N_2} (z_1,z_2)$ 
are analytic functions of $(z_1,z_2)$ and are given by
\beq
F_{N_1,N_2} ~\equiv ~\Big( {N! \over N_1! N_2!} \Big)^{1/2} ~z_1^{N_1} 
z_2^{N_2} ~.
\label{co2} 
\eeq
Eqs. (\ref{cs2}) and (\ref{co2}) are similar to (\ref{wcs}) and (\ref{co1}) 
respectively. This will be generalized to the $SU(3)$ case in section 3. 
It is easy to check that (\ref{cs2}) provides the resolution of identity 
with the measure given in (\ref{ints3}), namely,
\beq
\int ~d\Omega_{SU(2)} |z_1 , z_2 > < z_1 , z_2 | ~=~ \frac{1}{N+1} ~\sum_{m=
-j}^{j} ~|j,m><j,m| ~.
\eeq

\noindent Now we change variables from $N_1$ and $ N_2 = 2j - N_1$ to 
$m = {1 \over 2} (N_1 - N_2)$, and define 
\beq
\omega ~\equiv ~\frac{z_1}{z_2} ~=~ e^{i\phi} \cot \frac{\theta}{2} ~.
\eeq
These parameters are related to the ones given in (\ref{z1z2}) as
$\theta = 2 \chi$ and $\phi = \beta_1 - \beta_2$. We now
consider an unit sphere $S^2$ with its south pole touching the 
point $\omega=0$. The sphere is characterized by $(\theta,\phi)$ where 
$\theta$ and $\phi$ are the polar and azimuthal angles respectively. Using 
the stereographic projection, it is easy to verify that 
\bea
\vert z_1, z_2>_j & ~=~ & (z_1)^{2j} ~\sum_{m=-j}^j \sqrt{(2j)! \over (j+m)! 
(j-m)!} (\omega)^{(m-j)} \vert j, m > \nonumber \\
& =~ & \Big(z_1 cos(\frac{\theta}{2})\Big)^{2j} |\hat{n}(\theta,\phi)>_j ~, 
\label{equiv1} 
\eea
where we have again ignored possible phase factors. Eq. (\ref{equiv1})
can also be written as
\beq
\vert z_1, z_2>_j ~=~ (z_1)^{2j} \exp (\frac{z_2}{z_1} J_- ) \vert z_1 =1, 
z_2 =0> ~,
\label{equiv2}
\eeq
where $\vert z_1 =1, z_2 =0>_{N=2j} = \vert j,j>$ and we have used the fact
that $J_- = a_2^\dagger a_1$. Eqs. (\ref{equiv1}) and (\ref{equiv2}) establish 
the equivalence between the group theoretical theoretical definition 
(\ref{st}) and the one using Schwinger bosons (\ref{cs2}). 

The stationary subgroup of a particular coherent state is defined as the 
subgroup $H$ of the full group $G$ which leaves that coherent state invariant 
up to a phase; the coherent states are functions of the coset space $G/H$ 
\cite{perelomov}. It is clear from the discussion above that the stationary 
subgroup of the $SU(2)$ coherent states is $U(1)$; therefore the coherent 
states correspond to the coset space $SU(2)/U(1) = S^2$ which is parameterized
by the angles $(\theta ,\phi)$.

\section{\bf $SU(3)$ and its Representations} 

\indent
Let us first discuss a parameterization of $SU(3)$ matrices, i.e., $3 \times 
3$ unitary matrices with unit determinant. To motivate this, let us first
consider a parameterization of $SO(3)$ matrices. Consider a real vector of 
unit length of the form
\beq
{\vec p} ~=~ \left( \begin{array}{c} \sin \theta ~\cos \phi \\
\sin \theta ~\sin \phi \\
\cos \theta \end{array} \right) ~.
\eeq
The most general real vector $q$ of unit length which is orthogonal to $p$ 
is given by
\beq
{\vec q} ~=~ \left( \begin{array}{c} \cos \chi ~\cos \theta ~\cos \phi ~ +~ 
\sin \chi ~\sin \phi \\
\cos \chi ~\cos \theta ~\sin \phi ~-~ \sin \chi ~\cos \phi \\
- ~\cos \chi ~\sin \theta \end{array} \right) ~.
\eeq
Finally, we define a third unit vector ${\vec r} = {\vec p} \times {\vec q}$, 
i.e., $r_1 = p_2 q_3 - p_3 q_2$ etc. Then a $3 \times 3$ matrix whose columns 
are given by the vectors $p,q$ and $r$ is an $SO(3)$ matrix. 

We will now generalize the above construction to obtain an $SU(3)$ matrix.
A complex vector of unit norm is given by 
\beq
{\vec z} ~=~ \left( \begin{array}{c} \sin \theta ~\cos \phi ~e^{i\alpha_1} \\
\sin \theta ~\sin \phi ~e^{i\alpha_2} \\
\cos \theta ~e^{i\alpha_3} \end{array} \right) ~,
\label{defz}
\eeq
where $0 \le \theta , \phi \le \pi /2$ and $0 \le \alpha_1 , \alpha_2 , 
\alpha_3 < 2\pi$. 
Then the integration measure for $\vec z$, which is equivalent 
to the sphere $S^5$, is given by
\bea
d\Omega_{S^5} ~&=&~ \frac{2}{\pi^3} ~dz_1 ~d{\bar z}_1 ~dz_2 ~d{\bar z}_2 ~
dz_3 ~d{\bar z}_3 ~\delta (|z_1|^2+|z_2|^2+|z_3|^2-1) \nonumber \\
&=&~ \frac{1}{\pi^3} ~\sin^3 \theta ~\cos \theta ~\cos \phi ~
\sin \phi ~d \theta ~ d \phi ~d\alpha_1 ~d\alpha_2 ~d\alpha_3 ~,
\label{ints5}
\eea
which has been normalized to make $\int d\Omega_{S^5} =1$. The most general 
complex vector $\vec w$ of unit norm satisfying ${\vec z} \cdot {\vec w} =0$ 
is given by 
\beq
{\vec w} ~=~\left( \begin{array}{c} e^{i(\beta_1 -\alpha_1 )} ~\cos \chi ~\cos 
\theta ~ \cos \phi ~ +~ e^{i(\beta_2 - \alpha_1 )} ~ \sin \chi ~\sin \phi \\
e^{i(\beta_1 - \alpha_2)} ~\cos \chi ~\cos \theta ~\sin \phi ~-~ e^{i(\beta_2 
- \alpha_2)} ~\sin \chi ~\cos \phi \\
- ~e^{i(\beta_1 - \alpha_3)} ~\cos \chi ~\sin \theta \end{array} \right) ~,
\label{defw}
\eeq
where $0 \le \chi \le \pi /2$ and $0 \le \beta_1, \beta_2 < 2\pi$ just as
in the integration measure for $S^3$ in (\ref{ints3}).
We may now define a third complex vector of unit norm as ${\vec v} = 
{\vec {\bar z}} \times {\vec w}$, where ${\vec {\bar z}} \equiv {\vec 
{z^\star}}$. Then we can check that a $3 \times 3$ matrix whose columns 
are given by $z$, $\bar w$ and $v$, i.e.,
\beq
S ~=~ \left( \begin{array}{ccc} z_1 & {\bar w}_1 & {\bar z}_2 w_3 - {\bar z}_3
w_2 \\
z_2 & {\bar w}_2 & {\bar z}_3 w_1 - {\bar z}_1 w_3 \\
z_3 & {\bar w}_3 & {\bar z}_1 w_2 - {\bar z}_2 w_1 \end{array} \right) 
\label{su3mat}
\eeq
is an $SU(3)$ matrix. 

The integration measure for the group $SU(3)$ is given 
by a product of (\ref{ints5}) and (\ref{ints3}) as \cite{nmpc}
\beq
d\Omega_{SU(3)} ~=~ \frac{1}{2\pi^5} ~\sin^3 \theta ~\cos \theta ~\cos \phi ~
\sin \phi ~\cos \chi ~\sin \chi ~d\theta ~d\phi ~d\chi ~d\alpha_1 ~
d\alpha_2 ~d\alpha_3 ~d\beta_1 ~d\beta_2 ~,
\label{intsu3}
\eeq
which is normalized so that $\int d\Omega_{SU(3)}  =1$. To prove Eq. 
(\ref{intsu3}), we note that the matrix in (\ref{su3mat}) can be written as
a product of two $SU(3)$ matrices, i.e., $S = A_3 A_2$, where
\beq
A_3 ~=~ \left( \begin{array}{ccc} \sin \theta ~\cos \phi ~e^{i\alpha_1} & \cos 
\theta ~\cos \phi ~e^{i\alpha_1} & - \sin \phi ~e^{-i\alpha_2 -i\alpha_3} \\
\sin \theta ~\sin \phi ~e^{i\alpha_2} & \cos \theta ~\sin \phi ~e^{i\alpha_2} 
& \cos \phi ~e^{-i\alpha_1 -i\alpha_2} \\
\cos \theta ~e^{i\alpha_3} & - \sin \theta ~e^{i\alpha_3} & 0 \end{array} 
\right) ~,
\eeq
and 
\beq
A_2 ~=~ \left( \begin{array}{ccc} 1 & 0 & 0 \\
0 & \cos \chi ~e^{-i\beta_1} & \sin \chi ~e^{i\beta_2 - i \alpha_1 - i 
\alpha_2 - i \alpha_3} \\
0 & - \sin \chi ~e^{-i\beta_2 + i \alpha_1 + i \alpha_2 + 1 \alpha_3} & 
\cos \chi ~e^{i \beta_1} \end{array} \right) ~.
\eeq
The structure of the matrix $A_3$ is determined entirely by the 
three-dimensional complex vector which forms its first column; hence
the integration measure corresponding to it
is given by (\ref{ints5}). The matrix $A_2$ is determined by the 
two-dimensional complex vector which forms its second column; its contribution
to the integration measure is therefore given by (\ref{ints3}). Note that
although the parameter appearing in $A_2$ is $\beta_2 - \alpha_1 - \alpha_2 - 
\alpha_2$ instead of only $\beta_2$ as in (\ref{z1z2}), this makes no 
difference in the product measure given in
(\ref{intsu3}) since the differentials $d\alpha_i$ 
already appear in the integration measure coming from $A_3$. Incidentally, 
this procedure generalizes to any $SU(N)$; the integration measure is given by 
a product of measures for $S^{2N-1}$, $S^{2N-3}$, ..., $S^3$ \cite{nmpc}.

In short, we have defined two complex vectors ${\vec{z}}= (z_1,z_2,z_3)$ and 
${\vec{w}}=(w_1,w_2,w_3)$ in (\ref{defz}) and (\ref{defw}). These satisfy the 
constraints 
\bea
{\vec {\bar z}} \cdot {\vec z} ~=~ |z_1|^2 + |z_2|^2 + |z_3|^2 ~&=&~ 1~,
\nonumber \\
{\vec {\bar w}} \cdot {\vec w} ~=~ |w_1|^2 + |w_2|^2 + |w_3|^2 ~&=&~ 1~, 
\label{cons1} 
\eea
and
\beq
{\vec z} \cdot {\vec w} ~=~ z_1 w_1 + z_2 w_2 + z_3 w_3 ~=~ 0~. \nonumber \\
\label{cons2} 
\eeq
These constraints leave eight real degrees of freedom as required for $SU(3)$.
We will take $\vec z$ and $\vec w$ to transform respectively as the $3$ and 
$3^\star$ representation of $SU(3)$. Thus an $SU(3)$ transformation
acts on the matrix $S$ in Eq. (\ref{su3mat}) by multiplication from the left.

Let us now define two triplets of harmonic oscillator creation 
annihilation operators $(a_i,b_i)$, i=1,2,3, satisfying
\bea 
\big[a_i,a^{\dagger}_j \big] = & \delta_{ij} ~, ~~& \quad  
\big[b_i,b^{\dagger}_j \big] = \delta_{ij} ~, \nonumber  \\ 
\big[a_i,b_j\big] = & 0 ~,~~ & \quad \big[a_i,b^{\dagger}_j \big] = 0 ~.  
\eea
We will often denote these two triplets by $(\vec{a},\vec{b})$ 
and the two number operators by $N_a (\equiv \vec{a}^\dagger \cdot
\vec{a})$ and $N_b (\equiv \vec{b}^\dagger \cdot \vec{b})$. 
Similarly, their vacuum state is denoted by $|\vec{0}_a,\vec{0}_b>$. 
Henceforth, we will ignore the subscripts $a, b$ and will denote the 
vacuum state by $|\vec{0},\vec{0}>$, and the eigenvalues of $N_a$, $N_b$ 
by $N$ and $M$ respectively. 

Now let $\lambda^a$, $a=1,2,...,8$ be the generators of $SU(3)$ in the 
fundamental representation; they satisfy the $SU(3)$ Lie algebra $[\lambda^a ,
\lambda^b] = i f^{abc} \lambda^c$. Let us define the following operators
\beq
Q^a = a^\dagger \lambda^a a - b^\dagger \lambda^{*a} b ~,
\eeq
where $a^\dagger \lambda^a a \equiv a^{\dagger}_i \lambda^a_{ij} 
a_j$, and $b^\dagger \lambda^{*a} b \equiv b^{\dagger}_i 
\lambda^{*a}_{ij} b_j$. To be explicit,
\bea
Q^3 ~&=&~ \frac{1}{2} ~( a_1^\dagger a_1 ~-~ a_2^\dagger a_2 ~-~ b_1^\dagger 
b_1 ~+~ b_2^\dagger b_2 ) ~, \nonumber \\
Q^8 ~&=&~ \frac{1}{2 {\sqrt 3}} ~( a_1^\dagger a_1 ~+~ a_2^\dagger a_2 ~-~ 2 
a_3^\dagger a_3 ~-~ b_1^\dagger b_1 ~-~ b_2^\dagger b_2 ~+~ 2 b_3^\dagger 
b_3 ) ~, \nonumber \\
Q^1 ~&=&~ \frac{1}{2} ~( a_1^\dagger a_2 ~+~ a_2^\dagger a_1 ~-~ b_1^\dagger 
b_2 ~-~ b_2^\dagger b_1 ) ~, \nonumber \\
Q^2 ~&=&~ -\frac{i}{2} ~( a_1^\dagger a_2 ~-~ a_2^\dagger a_1 ~+~ b_1^\dagger 
b_2 ~-~ b_2^\dagger b_1 ) ~, \nonumber \\
Q^4 ~&=&~ \frac{1}{2} ~( a_1^\dagger a_3 ~+~ a_3^\dagger a_1 ~-~ b_1^\dagger 
b_3 ~-~ b_3^\dagger b_1 ) ~, \nonumber \\
Q^5 ~&=&~ -\frac{i}{2} ~( a_1^\dagger a_3 ~-~ a_3^\dagger a_1 ~+~ b_1^\dagger 
b_3 ~-~ b_3^\dagger b_1 ) ~, \nonumber \\
Q^6 ~&=&~ \frac{1}{2} ~( a_2^\dagger a_3 ~+~ a_3^\dagger a_2 ~-~ b_2^\dagger 
b_3 ~-~ b_3^\dagger b_2 ) ~, \nonumber \\
Q^7 ~&=&~ -\frac{i}{2} ~( a_2^\dagger a_3 ~-~ a_3^\dagger a_2 ~+~ b_2^\dagger 
b_3 ~-~ b_3^\dagger b_2 ) ~.
\label{su3}
\eea
It can be checked that these operators satisfy the $SU(3)$ algebra amongst 
themselves, i.e, $[Q^a,Q^b] = i f^{abc} Q^c$. Further, 
\bea
\big[Q^a, a^{\dagger}_i \big] ~&=&~ \lambda^a_{ji} a^{\dagger}_j ~, \quad 
\quad ~~ \big[Q^a, b^{\dagger}_i\big] ~ = - \lambda^{*a}_{ji} b^{\dagger}_j ~,
\nonumber \\
\hspace{1.0cm} 
\big[Q^a, a^\dagger \cdot a \big] ~&=&~~~  0 ~~~~, ~~ \quad \quad \big[Q^a, 
b^\dagger \cdot b\big] ~~~ = 0 ~~~, \nonumber \\
\hspace{1.0cm} \big[ Q^a, a^\dagger \cdot b^\dagger \big] ~&=&~~~ 0 ~~~~, ~~ 
\quad \quad \big[Q^a, a \cdot b \big] ~~~ = 0 ~~~. 
\label{cas} 
\eea

{}From Eqs. (\ref{cas}), it is clear that the three states $a^{\dagger}_i |
\vec{0},\vec{0}>$ with ($N=1,M=0$) and $b^{\dagger}_i |\vec{0},\vec{0}>$ with 
($N=0, M=1$)  transform respectively as the fundamental representation ($3$) 
and its conjugate representation ($3^\star$). By taking the direct product of 
$N$ ${\vec{a}}^\dagger$'s and M ${\vec{b}}^\dagger$'s we can now 
form higher representations. We now define an operator
\beq
O^{i_1i_2...i_N}_{j_1j_2...j_M} \equiv  
a^{\dagger}_{i_1} a^{\dagger}_{i_2} ... a^{\dagger}_{i_N}  b^{\dagger}_{j_1}
b^{\dagger}_{j_2} ... b^{\dagger}_{j_M} ~.  
\eeq
Under $SU(3)$ transformation the states defined as  $|\tilde{\psi}>_{(N,M)} 
\equiv O^{i_1i_2...i_N}_{j_1j_2...j_M} |\vec{0},\vec{0}>$ will all have $N_a = 
N$ and $N_b = M$, and will transform amongst themselves. Further, 
$ |\tilde{\psi}> = N |\tilde{\psi}>$ and $N_b |\tilde{\psi}> = M 
|\tilde{\psi}>$. However, these do not form an irreducible representation
because $\vec{a} \cdot \vec{b}$ and $\vec{a}^\dagger \cdot \vec{b}^\dagger$ 
are $SU(3)$ invariant operators (see (\ref{cas})). A general  
basis vector in the irreducible representation $(N,M)$ is obtained by 
subtracting the traces and completely symmetrizing in upper and lower indices 
\cite{georgi}. More explicitly, a state in $(N,M)$ representation is given by
\bea
&& |\psi > ^{i_1,i_2,...i_N}_{j_1,j_2,...,j_M} \equiv 
\Big[ O^{i_1i_2...i_N}_{j_1j_2...j_M} + L_1 \sum_{l_1=1}^N \sum_{k_1=1}^M 
\delta^{i_{l_1}}_{j_{k_1}} O^{i_1i_2..i_{l_1-1}i_{l_1 +1}..i_N}_{j_1j_2.. 
j_{k_1-1} j_{k_1 +1} ...j_M} 
\hspace{0.8cm} \nonumber \\
&& + L_2 \sum_{l_1,l_2=1}^N \sum_{k_1,k_2=1}^M 
\delta^{i_{l_1}}_{j_{k_1}} \delta^{i_{l_2}}_{j_{
k_2}} O^{i_1i_2..i_{l_1-1}i_{l_1+1}..i_{l_2-1}i_{l_2+1}..i_N}_{j_1j_2..
j_{k_1-1}j_{k_1+1}..j_{k_2-1}j_{k_2+1} ...j_M} \nonumber \\   
&& + L_3 \sum_{l_1,l_2,l_3=1}^N \sum_{k_1,k_2,k_3=1}^M \delta^{i_{l_1}}_{
j_{k_1}} \delta^{i_{l_2}}_{j_{k_2}} \delta^{i_{l_3}}_{j_{k_3}} 
O^{i_1i_2..i_{l_1-1}i_{l_1+1}..i_{l_2-1}i_{l_2+1}.. i_{l_3-1}i_{l_3
+1}.. i_N}_{j_1j_2..j_{k_1-1}j_{k_1+1}..
j_{k_2-1}j_{k_2+1}..j_{k_3-1}j_{k_3+1} ...j_M}~~+ ... \nonumber \\
&& + L_Q \sum_{l_1,l_2,l_3,..,l_Q=1}^N 
\sum_{k_1,k_2,k_3,..,k_Q=1}^{M} \delta^{i_{l_1}}_{j_{k_1}} 
\delta^{i_{l_2}}_{j_{k_2}} .. \delta^{i_{l_Q}}_{j_{k_Q}} O^{i_1i_2..
i_{l_1-1}i_{l_1+1}..i_{l_2-1}i_{l_2+1}.. i_{l_Q-1}i_{l_Q+1}.. i_N}_{j_1
j_2.. j_{k_1-1}j_{k_1+1}.. j_{k_2-1}j_{k_2+1}..j_{k_Q-1}j_{k_Q+1} ...
j_M} \Big] |\vec{0},\vec{0}>, \nonumber \\
&&
\label{bv} 
\eea
where $Q = {\rm Min} (N,M)$, 
\beq
L_q \equiv {(-1)^{q} ~(a^\dagger \cdot b^\dagger )^{q} \over {q! (N+M+1)
(N+M)(N+M-1)...(N+M+2-q})} ~,
\label{coef} 
\eeq
and all the sums in (\ref{bv}) are over different indices, i.e, $l_1 \neq 
l_2 ...\neq l_q$ and $k_1 \neq k_2 ...\neq k_q$. The coefficients in Eq. 
(\ref{coef}) are chosen to satisfy the tracelessness condition
\beq
\sum_{i_l,j_k=1}^{3} \delta^{i_l}_{j_k} |\psi > ^{i_1,i_2,...i_N}_{j_1, 
j_2,...,j_M} =  0, ~~ {\rm for ~all} ~~l=1,2...N, ~~{\rm and}~~ k=1,2...M ~.
\label{trace} 
\eeq

\noindent For future purposes, a more compact notation 
for describing all the states given above is to write 
\beq
O^{i_1i_2...i_N}_{j_1j_2...j_M} \equiv 
(a^{\dagger}_1)^{N_1} (a^{\dagger}_2)^{N_2} (a^{\dagger}_3)^{N_3} 
(b^{\dagger}_1)^{M_1} (b^{\dagger}_2)^{M_2} (b^{\dagger}_3)^{M_3} ~,
\label{onm}
\eeq
where $(N_i, M_i)$ denote all the possible eigenvalues of the occupation 
number operators $(a^{\dagger}_i a_i, b^{\dagger}_i b_i)$ satisfying
\beq
N_1 + N_2 + N_3 = N, \quad {\rm and} \quad M_1 + M_2 + M_3 = M ~.
\label{nm} 
\eeq
The action of (\ref{onm}) on the vacuum is given by
\beq
O^{N_1 N_2 N_3}_{M_1 M_2 M_3} ~\vert {\vec 0}, {\vec 0} > ~=~ (N_1 ! N_2 ! 
N_3 ! M_1 ! M_2 ! M_3 !)^{1/2} ~\vert {}^{N_1 N_2 N_3}_{M_1 M_2 M_3} > ~.
\label{nimi}
\eeq
We can now write the basis vectors of the representation $(N,M)$ as
\bea
|\psi > ^{i_1,i_2,...i_N}_{j_1,j_2,...,j_M} & \equiv &  
|\psi>^{N_1 N_2 N_3}_{M_1 M_2 M_3}  = \Bigl[ O^{N_1 N_2 N_3}_{M_1 M_2 M_3} 
+ \sum_{q=1}^Q L_q {\sum_{[\vec{\alpha}]_q}} ~ 
^{N_1}C_{\alpha_{1}} ~^{N_2} C_{\alpha_2} ~^{N_3}C_{\alpha_3} 
\nonumber \\ 
& & ^{M_1}C_{\alpha_{1}} ~^{M_2} C_{\alpha_2} ~^{M_3}C_{\alpha_3} 
\alpha_1! \alpha_2! \alpha_3! O^{N_{1}-\alpha_1 N_2-\alpha_2 
N_3-\alpha_3}_{M_1-\alpha_1 M_2-\alpha_2 M_3-\alpha_3} \Bigl] |{\vec 0},
{\vec 0} >.
\label{on}
\eea
In this equation, $[\vec{\alpha}]_q$ denotes the sets of 
three non-negative integers $(\alpha_1, \alpha_2, \alpha_3)$ satisfying 
$\alpha_1 + \alpha_2 + \alpha_3 = q$, and  $N_i - \alpha_i \ge 0 ~,~
M_i - \alpha_i \ge 0$ for $i=1,2,3$. The $\sum_{[\vec{\alpha}]_q}$ denotes 
a summation over all sets of three such integers. In the notation of 
Eq. (\ref{on}), the tracelessness condition (\ref{trace}) for the 
$(N+1,M+1)$ representation takes the form
\beq
\sum_{[\vec{\gamma}]_1} |\psi>^{N_1+\gamma_1 ~~ N_2+\gamma_2 ~~ N_3+
\gamma_3}_{M_1+\gamma_1 ~~ M_2+\gamma_2 ~~ M_3+\gamma_3} ~=~ 0 ~. 
\label{trace2} 
\eeq
The definition in (\ref{on}) satisfies the condition given in (\ref{trace2}). 
This can be verified by using the identity
\bea
&& \sum_{[\vec{\gamma}]_1} \sum_{[\vec{\alpha}]_q} \alpha_1 ! \alpha_2 ! 
\alpha_3 ! ^{N_1+\gamma_1} C_{\alpha_1}~ ^{N_2+\gamma_2} C_{\alpha_2} ~ ^{N_3
+\gamma_3} C_{\alpha_3} ~~ ^{M_1+\gamma_1} C_{\alpha_1}~ ^{M_2+\gamma_2} 
C_{\alpha_2} ~ ^{M_3+\gamma_3} C_{\alpha_3} \nonumber \\ 
&& ~~~~~~~~~ O^{N_1+\gamma_1-\alpha_1 N_2+\gamma_2-\alpha_2 N_3+\gamma_3-
\alpha_3}_{M_1+\gamma_1-\alpha_1 M_2+\gamma_2-\alpha_2 M_3+\gamma_3-\alpha_3} 
=\Big[ \Big( N+M+2-q \Big) \sum_{[\vec{\alpha}]_{q-1}} ~+~\big(\vec{a}^\dagger 
\cdot \vec{b}^\dagger \big) ~\sum_{[\vec{\alpha}]_q} \Big] \nonumber \\ 
&& ~~~~~~~~~~~~~~~~~~ \alpha_1 ! \alpha_2 ! \alpha_3 ! ~ ^{N_1} C_{
\alpha_1}~ ^{N_2} C_{\alpha_2} ~ ^{N_3} C_{\alpha_3} ~ ^{M_1} C_{
\alpha_1}~ ^{M_2} C_{\alpha_2}~ ^{M_3} C_{\alpha_3} O^{N_1-\alpha_1 N_2-
\alpha_2 N_3-\alpha_3}_{M_1-\alpha_1 M_2-\alpha_2 M_3-\alpha_3} ~.     
\nonumber \\
&&
\eea

The dimension $D(N,M)$ of the representation $(N,M)$ can be obtained as 
follows. For the $(N,0)$ representation, no tracelessness 
condition needs to be imposed, and the dimension is simply given by the
number of states in Eq. (\ref{nimi}) which satisfy $\sum_i N_i = N$ and
$\sum_i M_i =0$. This gives $D(N,0) = (N+1)(N+2)/2$. Similarly, 
$D(0,M)=(M+1)(M+2)/2$. Now $D(N,M)$ is given by the number of states
satisfying $\sum_i N_i = N$, $~\sum_i M_i =M$, which is equal to the product 
$D(N,0) D(0,M)$, {\it minus} the number of states satisfying $\sum_i N_i = 
N-1$, $~\sum_i M_i =M-1$, which is equal to $D(N-1,0) D(0,M-1)$; the 
subtraction is because of the tracelessness condition. This gives 
\beq
D(N,M) ~=~ \frac{1}{2} ~(N+1)(M+1)(N+M+2) ~.
\eeq

\section{\bf $SU(3)$ Coherent States} 

\indent
We now observe that the states in Eq. (\ref{bv}) can be extracted from the 
following generating function, 
\beq
|\vec{z},\vec{w}>_{(N,M)} ~\equiv ~{\sqrt {N!M!}} ~\exp ~(\vec{z} \cdot 
\vec{a}^\dagger ~+~ \vec{w} \cdot \vec{b}^\dagger ) ~|\vec{0},\vec{0}> ~,  
\label{coh1} 
\eeq
where we have to project onto the subspace of states with ${\vec a}^\dagger 
\cdot {\vec a} = N$ and ${\vec b}^\dagger \cdot {\vec b} = M$ to
obtain the representation $(N,M)$. More explicitly,
\bea
|\vec{z},\vec{w}>_{(N,M)} & ~=~& \frac{({\vec z} \cdot {\vec a}^\dagger)^N}{{
\sqrt {N!}}} ~\frac{({\vec w} \cdot {\vec b}^\dagger )^M}{{\sqrt {M!}}} ~|{
\vec 0}, {\vec 0}> \nonumber \\
&~ = ~& \sum_{N_1,N_2,N_3} \hspace{-0.4cm} {}^{\prime}~\sum_{M_1,M_2,
M_3} \hspace{-0.4cm} {}^{\prime} ~F_{\vec{N}, \vec{M}} \left(z_1,z_2,z_3;w_1,
w_2,w_3 \right) ~\vert {}^{N_1 N_2 N_3}_{M_1 M_2 M_3} > ~. \nonumber \\
&&
\label{coh2}
\eea
In (\ref{coh2}), $\sum^{\prime}$ implies that the occupation 
numbers $(N_i, M_i)$ satisfy Eq. (\ref{nm}), and $F_{\vec{N},\vec{M}} 
\left( \vec{z},\vec{w} \right)$ are given by
\bea
F_{\vec{N},\vec{M}} \left(\vec{z},\vec{w} \right) = 
\left( \frac{N!M!}{N_1 ! N_2 ! N_3 ! M_1 ! M_2 ! M_3 !} \right)^{1/2} 
~z_1^{N_1} z_2^{N_2} z_3^{N_3} w_1^{M_1} w_2^{M_2} w_3^{M_3} ~.
\label{co3} 
\eea 
On expanding the right hand side of (\ref{coh2}), the coefficients of 
$z_1^{N_1} z_2^{N_2} z_3^{N_3} w_1^{M_1} w_2^{M_2} w_3^{M_3}$ 
give the basis vectors of $SU(3)$ in the representation $(N,M)$. It is
important to note that the tracelessness conditions in Eq. (\ref{bv}) are
{\it automatically} satisfied by the state in (\ref{coh2}). This is because 
we can always replace $\vert {}^{N_1 N_2 N_3}_{M_1 M_2 M_3} > $ by the SU(3) 
basis vectors $ |\psi>^{N_1 N_2 N_3}_{M_1 M_2 M_3} $ defined in (\ref{on}). 

It is instructive to consider a specific example here. The coherent state of
the representation $(1,1)$, i.e., the adjoint representation of $SU(3)$, is
given by
\beq
|\vec{z},{\vec{w}}>_{(1,1)} ~=~ \sum_{i,j=1}^3 ~z_i w_j ~
a_i^\dagger b_j^\dagger ~|{\vec 0},{\vec 0}> ~. 
\label{rep11}
\eeq
We then see that the sum of the coefficients of the three states $\vert 
{}^{100}_{100} >$, $\vert {}^{010}_{010} >$ and $\vert {}^{001}_{001} >$ is 
zero due to the constraint in Eq. (\ref{cons2}). Hence there are only
eight linearly independent states on the right hand side of Eq. 
(\ref{rep11}) as there should be; these eight states can be taken to be
\bea
\vert V_1 > ~&=&~ \frac{1}{\sqrt 2} ~(\vert {}^{100}_{100}> ~-~ \vert 
{}^{010}_{010}> )~, \nonumber \\
\vert V_2 > ~&=&~ \frac{1}{\sqrt 6} ~(\vert {}^{100}_{100}> ~+~ \vert 
{}^{010}_{010}> ~-~ 2~\vert {}^{001}_{001}> )~, \nonumber \\
\vert V_3 > ~&=&~ \vert {}^{100}_{010}> ~, \quad \vert V_4 > ~=~ \vert 
{}^{010}_{100}> ~, \nonumber \\
\vert V_5 > ~&=&~ \vert {}^{100}_{001}>~, \quad \vert V_6 > ~=~ \vert 
{}^{001}_{100}> ~, \nonumber \\
\vert V_7 > ~&=&~ \vert {}^{010}_{001}> ~, \quad \vert V_8 > ~=~ \vert 
{}^{001}_{010}>~.
\label{rep11ex}
\eea

The states defined in Eq. (\ref{coh2}) will be called the coherent state
of the representation $(N,M)$. Note that the equations (\ref{nm}), 
(\ref{coh2}), (\ref{co3}) are analogous to the corresponding SU(2)
equations (\ref{con2}), (\ref{cs2}) and (\ref{co2}) respectively.   
The SU(3) coherent states (\ref{coh2}) are normalized to unity, i.e.,
\beq
{}_{(N,M)} < {\vec z}, {\vec w} \vert {\vec z}, {\vec w} >_{(N,M)} ~=~ 1 ~.
\label{norm}
\eeq
To prove this, we use the operator identities
\beq
e^A ~e^B ~=~ e^B ~e^A ~e^{[A,B]} ~, \quad {\rm and} \quad e^A ~B e^{-A} ~
=~ B ~+~ [A,B] ~, 
\label{iden}
\eeq
which hold if $[A,B]$ commutes with both $A$ and $B$. We find that 
\beq
<{\vec 0}, {\vec 0} | ~\exp ~[{\vec {\bar z}} \cdot {\vec a}+ {\vec {\bar w}}
\cdot {\vec b}] ~\exp ~[{\vec z} \cdot {\vec a}^\dagger + {\vec w} \cdot 
{\vec b}^\dagger ] ~| {\vec 0}, {\vec 0} > ~=~ \exp ~[ {\vec {\bar z}} \cdot 
{\vec z} ~+~ {\vec {\bar w}} \cdot {\vec w} ] ~.
\eeq
On comparing terms of order $({\vec {\bar z}} \cdot {\vec z})^N ({\vec {\bar 
w}} \cdot {\vec w})^M$ on both sides of this equation and using the
definition in (\ref{coh2}), we obtain Eq. (\ref{norm}). In the same way,
we can show that 
\beq
{}_{(N,M)}< {\vec z}, {\vec w} \vert {\vec z}+d{\vec z}, {\vec w}+d{\vec w} 
>_{(N,M)} ~=~ 1 ~+~ N ~\sum_i ~{\bar z}_i dz_i ~+~ M ~\sum_i ~{\bar w}_i 
dw_i ~,
\label{diff}
\eeq
where $d{\vec z}$ and $d{\vec w}$ denote small deviations from $\vec z$
and $\vec w$. This equation will be used to derive the path integral
formalism \cite{manous,fradkin} in section 5, and it would also be useful for 
obtaining the geometric phase for systems with $SU(3)$ symmetry \cite{khanna}. 
 
We can prove that the states defined in Eq. (\ref{coh2}) satisfy the 
resolution of identity, i.e, 
\beq
\int d\Omega ~|\vec{z},\vec{w}>_{(N,M)} {}_{(N,M)}<\vec{z},\vec{w}| 
~=~ \frac{1}{D(N,M)} ~\sum_{i=1}^{D(N,M)} ~\vert V_i > < V_i \vert ~,
\label{roi2} 
\eeq
where $V_i$ denotes a set of 
orthonormal basis vectors of $(N,M)$. (See Eq. (\ref{rep11ex}) for the 
explicit example of the representation $(1,1)$). To verify the normalization
on the right hand side of Eq. (\ref{roi2}), it is convenient to look at 
a particular basis vector $\vert {}^{N00}_{0M0} >$. (This has the maximum 
eigenvalue $(N+M)/2$ of the operator $Q^3$ given in Eq. (\ref{su3})). From 
Eq. (\ref{coh2}), the coefficient of this vector in the coherent state 
is given by $z_1^N w_2^M$. Integrating the modulus squared of 
this using Eqs. (\ref{defz} - \ref{intsu3}), we obtain the factor of
$1/D(N,M)$ in Eq. (\ref{roi2}). This is as it should be so that taking the
trace of both sides of (\ref{roi2}) gives unity.

A second property of coherent states is that they are overcomplete. This is
clear for the states in (\ref{coh2}) since they are continuous 
functions of the complex variables $({\vec z},{\vec w})$, while the 
dimension of the representation $(N,M)$ is finite.

The coherent states in (\ref{coh2}) have a third property which is group
theoretical, and is analogous to Eq. (\ref{equiv2}) for the $SU(2)$ coherent 
states. Namely, we can go from a particular coherent state, say, $| z_1 
=1, w_2 = 1>_{(N,M)} = \vert {}^{N00}_{0M0}>$ to the general coherent state 
$|z, w>_{(N,M)}$by acting with an exponential of certain combinations
of the $SU(3)$ generators $Q^a$. First of all, we can check that 
\beq
|z, w>_{(N,M)} = z_1^N w_2^M ~\exp ~\left[ \frac{z_2}{z_1} ~a_2^\dagger 
a_1 + \frac{z_3}{z_1} ~a_3^\dagger a_1 + \frac{w_1}{w_2} ~b_1^\dagger 
b_2 + \frac{w_3}{w_2} ~b_3^\dagger b_1 \right]~ | z_1 =1, w_2 =1>_{(N,M)} ~.
\eeq
Then we can use Eq. (\ref{iden}) and the constraint (\ref{cons2}) to rewrite 
this in the form \cite{gnutzmann}
\bea
&& |z, w>_{(N,M)} \nonumber \\
&& = z_1^N w_2^M ~\exp ~\left[ \frac{z_2}{z_1} ~(Q^1 - iQ^2)
+ \frac{z_3}{z_1} ~(Q^4 - iQ^5) - \frac{w_3}{w_2} ~(Q^6 + iQ^7) \right]~ | 
z_1 =1, w_2 =1>_{(N,M)} , \nonumber \\
&&
\label{group}
\eea
which is similar in structure to Eq. (\ref{equiv2}).

Another property of these coherent states which is important for their path 
integral applications is that the expectation value of the $SU(3)$ operators 
(\ref{su3}) in a coherent state should be given by an $SU(3)$ covariant 
function of $({\vec z},{\vec w})$ and their complex conjugates. We find that
\beq
{}_{(N,M)}<\vec{z},\vec{w}| Q^a |\vec{z},\vec{w}>_{(N,M)} ~=~
N ~{\bar z}_i \lambda^a_{ij} z_j ~-~ M ~{\bar w}_i \lambda^{*a}_{ij} w_j ~.
\label{expec1}
\eeq
This can be proved by using the identities in Eq. (\ref{iden}) to show that
\beq
<{\vec 0}, {\vec 0} | ~\exp ~[{\vec {\bar z}} \cdot {\vec a} + {\vec {\bar w}}
\cdot {\vec b}] ~a_i^\dagger ~a_j ~\exp ~[{\vec z} \cdot {\vec a}^\dagger + 
{\vec w} \cdot {\vec b}^\dagger ] ~| {\vec 0}, {\vec 0} > ~=~
{\bar z}_i z_j ~\exp ~[{\vec {\bar z}} \cdot {\vec z} ~+~ {\vec {\bar w}} 
\cdot {\vec w}] ~,
\label{expec2}
\eeq
and a similar identity for the expectation value of $b_i^\dagger b_j$ in terms 
of ${\bar w}_i w_j$. Eq. (\ref{expec1}) can now be obtained by comparing terms 
of order ${\bar z}^N z^N {\bar w}^M w^M$ on the two sides of Eq. 
(\ref{expec2}).

The stationary subgroup of the coherent states defined in this section is
generally 
$U(1) \times U(1)$, corresponding to multiplying the vectors $\vec z$ and
$\vec w$ by independent phase factors. These coherent states are therefore 
functions of the coset space $SU(3)/U(1) \times U(1)$ \cite{gnutzmann}.
However, for the completely symmetric representations $(N,0)$ and $(0,M)$,
the coherent states use only three complex numbers ($\vec z$ or $\vec w$)
which define the space $S^5$; the stationary subgroup is then $U(1) = S^1$
which corresponds to multiplying that complex vector by a phase factor. In
those cases, the coherent states are functions of the manifold $S^5 /S^1$.

\section{\bf An Alternative Definition of $SU(3)$ Coherent States} 

\indent
The $SU(3)$ coherent states discussed in section 4 involve eight real 
parameters, and satisfy some simple group theoretic properties similar to the 
$SU(2)$ coherent states of section 2. It is possible that there may be some 
applications of coherent states which do not require so many parameters. In 
this section, we will discuss an alternative kind of coherent states which 
only require five real parameters. We will see later that these coherent 
states suffer from some problems and they seem to lack some of the group 
theoretic properties precisely because they use fewer parameters. 

We observe that the states in (\ref{bv}) can be extracted 
from the following generating function
\beq
|\vec{z},\vec{\bar{z}}> \equiv \exp ~(\vec{z} \cdot \vec{a}^\dagger) ~
\exp ~(\vec{\bar{z}} \cdot \vec{b}^\dagger ) ~\Big[ ~1 + \sum_{q=1}^{Q} L_q ~
\Big] ~|\vec{0},\vec{0}> ~,  
\eeq
and we have to project onto the subspace of states with ${\vec a}^\dagger 
\cdot {\vec a} = N$ and ${\vec b}^\dagger \cdot {\vec b} = M$ to obtain the 
representation $(N,M)$. To be explicit,
\beq
|\vec{z},{\vec{\bar{z}}}>_{(N,M)} ~=~ \Bigl[ \frac{({\vec z} \cdot {\vec 
a}^\dagger )^N}{N!} ~\frac{({\vec {\bar z}} \cdot {\vec b}^\dagger )^M}{
M!} ~+~ \sum_{q=1}^Q ~L_q ~\frac{({\vec z} \cdot {\vec a}^\dagger )^{N-q}}{(
N-q)!} ~\frac{({\vec {\bar z}} \cdot {\vec b}^\dagger )^{M-q}}{(M-q)!}~ 
\Bigl] |{\vec 0}, {\vec 0}> ~.
\label{co4}
\eeq
On expanding the right hand side of (\ref{co4}), the coefficients of the 
tensors $z_{i_1} z_{i_2} ... z_{i_N} \bar{z}_{j_1} \bar{z}_{j_2}... \bar{
z}_{j_M}$ give the basis vectors of $SU(3)$ in the representation $(N,M)$. 

The $SU(3)$ coherent states in the representation $(N,M)$ are defined as 
in Eq. (\ref{co4}),
\bea
|\vec{z},{\vec{\bar{z}}}>_{(N,M)} &\equiv & \frac{1}{N!M!} ~
\sum_{i_1,i_2,...} \sum_{j_1,j_2,...} z_{i_1} z_{i_2} ... z_{i_N} 
\bar{z}_{j_1} \bar{z}_{j_2} ... \bar{z}_{j_M} |\psi>^{i_1i_2...i_N}_{j_1
j_2...j_M} \nonumber \\
&=& \sum_{N_1, N_2, N_3} \sum_{M_1, M_2, M_3} ~{z_1^{N_1} z_2^{N_2} z_3^{N_3} 
{\bar z}_1^{M_1} {\bar z}_2^{M_2} {\bar z}_3^{M_3} \over {N_1}! {N_2}! 
{N_3}! {M_1}! {M_2}! {M_3}!} ~|\psi>^{N_1 N_2 N_3}_{M_1 M_2 M_3} ~.
\label{cs3} 
\eea
To give a specific example, the coherent state of the representation $(1,1)$ 
is given by
\beq
|\vec{z},{\vec{\bar{z}}}>_{(1,1)} ~=~ \sum_{i,j=1}^3 ~z_i {\bar z}_j 
a_i^\dagger b_j^\dagger |{\vec 0},{\vec 0}> ~-~ \frac{1}{3} ~ 
\sum_{i=1}^3 ~a_i^\dagger b_i^\dagger |{\vec 0}, {\vec 0}> ~.
\eeq

We will now prove that the states defined in (\ref{cs3}) satisfy the 
resolution of identity, 
\bea
\int ~d\Omega_{S^5} ~|\vec{z},\vec{\bar{z}}>_{(N,M)} {}_{(N,M)}<\vec{z},
\vec{\bar{z}}| ~=~ 1 ~. 
\eea
To prove this, we use the the definition (\ref{on}) and the integration 
measure for $\vec z$ given in (\ref{ints5}). We find that 
\bea
&& \int d\Omega_{S^5} ~|z,\bar{z}> <z,\bar{z}| \nonumber \\
&& =~  {\cal C} \sum_{N_i ,M_i} \Big( \sum_{\delta_i} \big( \prod_{i=1}^3 
{(N_i+M_i+\delta_i)! \over (N_i+\delta_i)! (M_i+\delta_i)!} \big) ~
|\psi>^{N_1+\delta_1 N_2+\delta_2 N_3+\delta_3}_{M_1+\delta_1 
M_2+\delta_2 M_3+\delta_3} \Big) ~{}^{N_1 N_2 N_3}_{M_1 M_2 M_3}< \psi| ~,
\label{xy1}
\eea
where the $\delta_i$ are integers satisfying 
\beq
\sum_{i=1}^3 ~\delta_i ~=~ 0 ~,
\eeq
and the constant $\cal C$ is determined below. We now use the following 
property
\beq
\sum_{\delta_i} \big( \prod_{i=1}^3 
{(N_i+M_i+\delta_i)! \over (N_i+\delta_i)! (M_i+\delta_i)!} \big) ~
|\psi> ^{N_1+\delta_1 N_2+\delta_2 N_3+\delta_3}_{M_1+\delta_1 
M_2+ \delta_2 M_3+\delta_3} = |\psi> ^{N_1 N_2 N_3}_{M_1 M_2 M_3} ~,
\eeq
which is a consequence of Eq. (\ref{trace}) for the basis vectors of a 
representation of $SU(3)$. Thus Eq. (\ref{xy1}) can be simplified to
\beq
\int d\Omega_{S^5} ~|z,\bar{z}> <z,\bar{z}| ~=~ {\cal C} \sum_{N_i, M_i} 
|\psi>^{N_1 N_2 N_3}_{M_1 M_2 M_3} ~{}^{N_1 N_2 N_3}_{M_1 M_2 M_3}< \psi| ~.
\label{xy2} 
\eeq
The normalization constant $\cal C$ in Eq. (\ref{xy2}) can be fixed by looking
at one particular basis vector of the representation $(N,M)$, say,
\beq
| \psi >^{N00}_{0M0} ~.
\eeq
{}From Eq. (\ref{cs3}), the coefficient of this vector in the coherent state 
$|{\vec z},{\vec {\bar z}}>$ is $z_1^N {\bar z}_2^M /(N!M!)$. Integrating 
this as in (\ref{ints5}), we find that
\beq
{\cal C} ~=~ \frac{2}{N! M! (N+M+2)!} ~.
\eeq

Finally, let us consider the analog of the property given in Eq. 
(\ref{expec1}) for the $(z,w)$ coherent states. We can prove that
\beq
{}_{(N,M)}<\vec{z},\vec{\bar{z}}| Q^a |\vec{z},\vec{\bar{z}}>_{(N,M)} ~=~
(N-M) ~{\bar z}_i \lambda^a_{ij} z_j ~.
\label{expec3}
\eeq
To prove this, we use the identities in (\ref{iden}) to show that 
\beq
<{\vec 0}, {\vec 0} | ~\exp ~[{\vec {\bar z}} \cdot {\vec a} + {\vec z}
\cdot {\vec b}] ~a_i^\dagger ~a_j ~\exp ~[{\vec z} \cdot {\vec a}^\dagger + 
{\vec {\bar z}} \cdot {\vec b}^\dagger ] ~| {\vec 0}, {\vec 0} > ~=~
{\bar z}_i z_j ~\exp ~[2 {\vec {\bar z}} \cdot {\vec z}] ~.
\eeq
On expanding this equation and comparing terms which are
of order $N$ in both $z_i$ and ${\bar z}_i$, we 
find that the expectation value of $Q^a$ in the representation $(N,0)$ 
satisfies Eq. (\ref{expec3}). In a similar way, we can
prove Eq. (\ref{expec3}) in the representation $(0,M)$. Finally, we can
generalize the proof to the representation $(N,M)$ by using Eq. (\ref{cas});
since $Q^a$ commutes with ${\vec a} \cdot {\vec b}$ and ${\vec a}^\dagger 
\cdot {\vec b}^\dagger$, it also commutes with the operators $L_q$ 
which are require to enforce tracelessness in Eq. (\ref{bv}).

Note that (\ref{expec3}) vanishes for the self-conjugate representations 
in which $N=M$. There is a similar problem for the differential change in 
overlap analogous to Eq. (\ref{diff}). We find that the coherent states 
defined in this section satisfy
\beq
\frac{<{\vec z}, {\vec {\bar z}}\vert {\vec z} + d{\vec z}, {\vec {\bar z}}
+ d {\vec {\bar z}}>}{<{\vec z}, {\vec {\bar z}}\vert {\vec z}, {\vec {\bar 
z}}>} ~=~ 1 ~+~ N ~\sum_i ~{\bar z}_i dz_i ~+~ M ~\sum_i d{\bar z}_i z_i 
\eeq
in the representation $(N,M)$. The left hand side of this equation is equal 
to $1$ if $N=M$ due to the constraint $\sum_i {\bar z}_i z_i =1$. These two 
problems imply that the $(z,{\bar z})$ coherent states are unlikely to be 
useful for path integral applications in the representations with $N=M$. 

For the $(z,{\bar z})$ coherent states, we have not yet found the construction 
of the group theoretical property analogous to (\ref{group}) in the general 
representation $(N,M)$. This would be an interesting topic for future studies.

The stationary subgroup of the coherent states defined in this section is
$U(1) = S^1$, corresponding to multiplying $\vec z$ by a phase factor. These 
coherent states are therefore functions of the manifold $S^5 /S^1$.

\section{\bf Path Integral Formalism}

\indent
We will now use the $(z,w)$ coherent states presented in section 4 to derive 
the path integral for a problem which has $SU(3)$ variables in some 
representation $(N,M)$. (For convenience, we will drop the subscript $(N,M)$ 
on the coherent states in this section). We begin by discussing a problem 
involving the Hamiltonian of a single site with a $SU(3)$ variable. For any
Hamiltonian which is a function of the $SU(3)$ operators $Q^a$, we define
its coherent state expectation value to be
\beq
E (z,{\bar z},w,{\bar w}) ~\equiv ~ <z,w| {\hat H} |z,w> ~.
\label{ezw}
\eeq
If the Hamiltonian is linear in the $SU(3)$ operators, i.e., 
\beq
{\hat H} ~=~ \sum_{a=1}^8 ~c_a Q^a ~,
\eeq
then Eq. (\ref{ezw}) can be found found using Eq. (\ref{expec1}). But if the
Hamiltonian is not linear in the $SU(3)$ operators, then Eq. (\ref{ezw})
has to be evaluated separately. 

Let us now consider the propagator in imaginary time 
\beq
G (z^{(F)},w^{(F)},z^{(I)},w^{(I)};T) ~=~ <z^{(F)},w^{(F)} | \exp (- T 
{\hat H}) |z^{(I)},w^{(I)} > ~,
\label{prop}
\eeq
where the superscripts $I$ and $F$ denote initial and final states 
respectively, and we are suppressing the subscripts $i$ ($=1,2,3$) on $z$ and 
$w$ for the moment. We write the exponential in (\ref{prop}) as a product of 
$\cal N$ terms, and use the resolution of identity in (\ref{roi2}) to insert a 
complete set of states between each pair of terms. A typical term looks like
\beq
<z^{(n+1)},w^{(n+1)}| \exp (-\epsilon {\hat H}) |z^{(n)},w^{(n)} > ~, 
\label{path1}
\eeq
where $\epsilon = T/{\cal N}$. We are eventually interested in taking the limit
${\cal N} \rightarrow \infty$ holding $T$ fixed. In that case, we may assume
that $(z^{(n+1)},w^{(n+1)})$ is close to $(z^{(n)},w^{(n)})$ in (\ref{path1}), 
so that $dz^{(n)}_i = z^{(n+1)}_i - z^{(n)}_i$ and $dw^{(n)}_i = w^{(n+1)}_i - 
w^{(n)}_i$ are small. Using Eqs. (\ref{diff}) and (\ref{ezw}), we can write 
(\ref{path1}) as
\bea
&& <z^{(n+1)},w^{(n+1)}| \exp (-\epsilon {\hat H}) |z^{(n)},w^{(n)} > 
\nonumber \\
&& =~ \exp [N \sum_i {\bar z}^{(n)}_i dz^{(n)}_i ~+~ M \sum_i {\bar w}^{(n)}_i 
dw^{(n)}_i ~-~ \epsilon E (z^{(n)}, {\bar z}^{(n)}, w^{(n)}, {\bar w}^{(n)})] 
\eea
to first order in $\epsilon$, $dz^{(n)}_i$ and $dw^{(n)}_i$. In the limit 
$\epsilon = d\tau \rightarrow 0$, we can write the propagator in (\ref{prop}) 
in the path integral form
\bea
G (z^{(F)},w^{(F)},z^{(I)},w^{(I)};T) &=& \int ~{\cal D} \Omega_{SU(3)} 
(\tau) ~\exp (-S [z,w]) ~, \nonumber \\
{\rm where} \quad S [z,w] &=& \int_0^T ~d\tau ~[-~ N \sum_i {\bar z}_i 
\frac{dz_i}{d\tau} ~-~ M \sum_i {\bar w}_i \frac{dw_i}{d\tau} ~+~ 
E(z,{\bar z},w,{\bar w}) ]~, \nonumber \\
{\rm and} \quad {\cal D} \Omega_{SU(3)} (\tau) &\equiv & \prod_n ~
d\Omega_{SU(3)} (n) ~,
\label{path2}
\eea
and $(z,w)$ are functions of $\tau$ which satisfy the boundary conditions $(
z(0),w(0)) = (z^{(I)}, w^{(I)})$ and $(z(T),w(T)) = (z^{(F)}, w^{(F)})$. Note 
that we have written the functional integral measure in (\ref{path2}) in terms 
of the measure given in Eq. (\ref{intsu3}). Alternatively, we can write the 
functional integral measure in terms of ${\cal D} z {\cal D} {\bar z}{\cal D} 
w {\cal D} {\bar w}$ if we introduces appropriate Lagrange multiplier fields 
in the action $S$ to enforce the constraints in Eqs. (\ref{cons1} - 
\ref{cons2}) at each time $\tau$.

We can now generalize the above construction to a problem involving several
sites which are labelled by a parameter $x$, provided that the Hamiltonian 
is linear in the $SU(3)$ variables at {\it each} site. We introduce a 
coherent state at each site, and write the energy functional as
\bea
E [z, {\bar z}, w, {\bar w}] ~&=&~ <z,w |{\hat H}| z, w > ~,
\nonumber \\
{\rm where} \quad |z, w> ~& \equiv &~ \prod_x ~| z(x), w(x)> ~.
\eea
Then we can show that 
\bea
&& < z^{(F)} (x), w^{(F)} (x) | \exp (-T {\hat H}) | z^{(I)} (x), w^{(I)} 
(x) > ~= \int ~{\cal D} \Omega_{SU(3)} (x,\tau ) ~\exp (- S [z,w]) ~, 
\nonumber \\
&& S [z,w] ~= \int_0^T ~d\tau ~\left[ ~- \sum_x ~\{ ~ N \sum_i {\bar z}_i (x) 
\frac{dz_i (x)}{d \tau} ~-~ M \sum_i {\bar w}_i (x) \frac{dw_i (x) }{d\tau} ~
\} ~+~ E [z, {\bar z}, w, {\bar w} ] \right] , \nonumber \\
&& {\cal D} \Omega_{SU(3)} (x,\tau ) \equiv \prod_{x,n} ~d\Omega_{SU(3)} 
(x,n) ~.
\label{path3}
\eea
Note that the first two terms in the actions $S$ given in Eqs. (\ref{path2}) 
and (\ref{path3}) are purely imaginary due to the constraints in (\ref{cons1}). 
To show this explicitly, we can rewrite those terms as
\bea
\sum_i ~{\bar z}_i dz_i ~&=&~ \frac{1}{2} ~\sum_i ~(~ {\bar z}_i dz_i ~-~ 
d{\bar z}_i z_i ~)~, \nonumber \\
{\rm and} \quad \sum_i ~{\bar w}_i dw_i ~&=&~ \frac{1}{2} ~\sum_i ~(~ 
{\bar w}_i dw_i ~-~ d{\bar w}_i w_i ~)~.
\eea

As an example of a problem to which this formalism can be applied, we can
consider the $SU(3)$ invariant Hamiltonian
\beq
{\hat H} ~=~ \sum_{x,y} ~J_{x,y} ~\sum_a ~Q^a (x) Q^a (y) ~.
\eeq
This is called the $SU(3)$ Heisenberg model. It has been discussed extensively
in the literature for the completely symmetric representations $(N,0)$
\cite{manous}; for those representations, we can use the simpler measure
$d\Omega_{S^5}$ given in Eq. (\ref{ints5}) instead of $d\Omega_{SU(3)}$. Our 
construction of coherent states now allows a study of the Heisenberg model 
in any representation $(N,M)$. 

\section{\bf Summary and Discussion} 

\indent
In this paper we have exploited the representation of the $SU(3)$ Lie algebra  
in terms of six harmonic oscillator creation and annihilation 
operators to generate all the representations of $SU(3)$. This harmonic 
oscillator form of the algebra enables us to define the $SU(3)$ 
coherent states in terms of two triplets of complex numbers. In this sense 
the $SU(2)$ (\ref{cs2}) and $SU(3)$ definitions (\ref{coh1}) are analogous 
to that of the Heisenberg-Weyl coherent states (\ref{wcs}). The
$SU(3)$ coherent states are characterized by two triplets of complex 
numbers with $4$ real constraints. This explicit construction in terms 
of complex numbers can be used to derive the geometrical phase of $SU(3)$. 
Further, the path integral formalism discussed in the previous section can be 
used to obtain the field theory for the $SU(3)$ Heisenberg model and 
study its topological aspects as in the $SU(2)$ case \cite{haldane}. Work in 
this direction is in progress and will be reported elsewhere.   

For any group $G$, we can use a certain number of harmonic 
oscillator operators to construct the group operators as in Eqs. (\ref{su3}) 
and (\ref{cas}). If we can find the appropriate set of complex numbers
which transform according to that group and satisfy the necessary 
constraints, we can use our method to provide an explicit complex number 
parameterization of the corresponding coherent states.  

\vskip .8 true cm

\leftline{\bf Acknowledgments}

We  would like to thank N. Mukunda and H. S. 
Sharatchandra for  discussions. M.M. would like to thank Samir Paul,
Debashish Gangopadhyay  and Ranjan Choudhary for discussions 
on the $SU(2)$ coherent states.

\end{document}